\documentclass[aps,prl,showpacs,twocolumn]{revtex4}

\usepackage{amsmath,epsfig,amsfonts,epsfig,graphicx,
}

\begin{document}
\title{Nonlinearity Management in Optics: Experiment, Theory, and Simulation}
\author{Martin Centurion$^{1,2}$, Mason A. Porter$^{2,3}$, P. G. Kevrekidis$^4$, and Demetri
Psaltis$^{1}$ }
\affiliation{ $^1$Department of Electrical Engineering,
$^2$Center for the Physics of Information,
$^3$Department of Physics, California Institute of Technology, Pasadena, CA 91125, USA \\
$^4$Department of Mathematics and Statistics, University of Massachusetts, Amherst MA 01003-4515, USA }

\begin{abstract}

We conduct an experimental investigation of nonlinearity management in optics using femtosecond pulses and layered Kerr media consisting of glass and air.  By examining the propagation properties over several diffraction lengths, we show that wave collapse can be prevented.  We corroborate these experimental results with numerical simulations of the (2 + 1)-dimensional focusing cubic nonlinear Schr\"odinger equation with piecewise constant coefficients and a theoretical analysis of this setting using a moment method.

\end{abstract}


\pacs{05.45.Yv, 42.65.Sf, 42.65.Tg, 42.65.-k}

\maketitle


{\it Introduction}. In the past decade, techniques for managing dispersion \cite{DM} and nonlinearity \cite{borisbook} have attracted considerable
attention in diverse branches of physics, including optics, atomic/condensed-matter physics, and other areas. Dispersion management (DM), originally proposed for optical fibers based on periodically alternating the group-velocity dispersion with opposite signs, was found to be a robust method for supporting breathing solitary waves \cite{DMS}.  It was subsequently implemented in atomic matter waves \cite{ober}, where the sign of dispersion was controlled via periodic potentials.
Nonlinearity management (NM) was originally proposed in the
context of layered optical media \cite{mezentsev}, but it has garnered special attention in the context of ultracold physics \cite{borisbook}, where it was reformulated as Feshbach Resonance Management \cite{frm}. In this latter setting, the Feshbach resonance was proposed as a means of periodically modifying the scattering length of interatomic interactions (and hence the mean-field nonlinearity)
to avoid collapse in higher-dimensional Bose-Einstein condensates \cite{saito} and sustain robust breathing coherent structures of atomic matter waves \cite{montesinos,others}.  The idea of NM has
also inspired considerable mathematical research,
with a focus on its averaged and collapse-preventing properties \cite{pelinovsky}.


However, to the best of our knowledge, there has not yet been any {\it experiment} that addresses the theoretically proposed framework of NM. The present Letter aims to close this gap by offering an experimental investigation of NM in an optical setting.  In particular, we study the propagation of femtosecond pulses in layered Kerr media.  In a nonlinear Kerr medium, the index of refraction can be written $n = n_0 + n_2I$, where $n_0$ is the linear contribution, $n_2$ is the Kerr coefficient, and $I$ is the beam intensity.  A beam propagating through such a medium with self-focusing (i.e., $n_2>0$)
in two-dimensions (2D)
will
collapse \cite{sulem} if the beam power is above a critical threshold,
$P_c = \pi(0.61)^2\lambda^2/(8n_0n_2)$,
where $\lambda$ is the light beam's wavelength.
 Our medium has layers of
 glass and air, which are both self-focusing but with very different Kerr coefficients ($n_2^{(1)} = 3.2 \times 10^{-16} \rm{cm}^2/\rm{W}$ for glass \cite{Tzortzakis} and $n_2^{(2)} = 3.2 \times 10^{-19} \rm{cm}^2/\rm{W}$ for air \cite{Couairon}). The
 aim of our experiment is to showcase what is arguably NM's most striking feature---preventing the wave collapse in 2D that would
occur in a homogeneous medium.  We illustrate this effect for powers in the interval ($2 P_c, 6P_c$), where $P_c$ refers to the critical power in glass, that would otherwise be on the verge of collapse before the occurrence of more complex processes, such as the formation of a plasma through multiphoton absorption \cite{Braun, Liu}. The latter offers a defocusing, lossy mechanism that prevents this catastrophic phenomenon.  Optical beams that do not diverge due to diffraction (filaments or spatial solitons) have been obtained through a variety of physical mechanisms \cite{Segev}. Self-guiding and filamentation of femtosecond pulses
relies on nonlinear losses and negative index changes from plasma formation to stabilize the beam \cite{Tzortzakis,Couairon,Braun,Liu,Dubietis}. Here we demonstrate a new guiding mechanism that does not rely on plasma formation
and can, in principle, be lossless. The variation of the Kerr coefficient in the layered medium prevents collapse, sustaining an oscillatory variation of the beam width for considerable propagation distances (before eventual dispersion due to weak losses arising at the interfaces between air-glass slides). We
 compare our experimental results to numerical simulations of the nonlinear Schr{\"o}dinger (NLS) equation,
 adapted to the detailed experiment setup, and find good qualitative agreement (and quantitative agreement within the appropriate propagation regime) between the two. An additional theoretical understanding of the experimental trends is offered by a moment approach \cite{montesinos}.


\begin{figure}[tbp]
\centering \includegraphics[width=7.5cm]{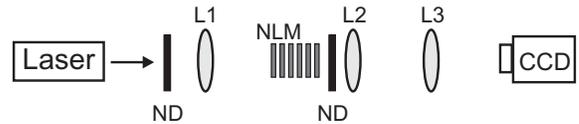}
\caption{Experimental setup. ND = Neutral density filters, NLM =
Nonlinear medium, and L1--L3 = lenses.} \label{setup}
\end{figure}

\begin{figure}[tbp]
\centering \includegraphics[height=5.75cm,width=7.5cm]{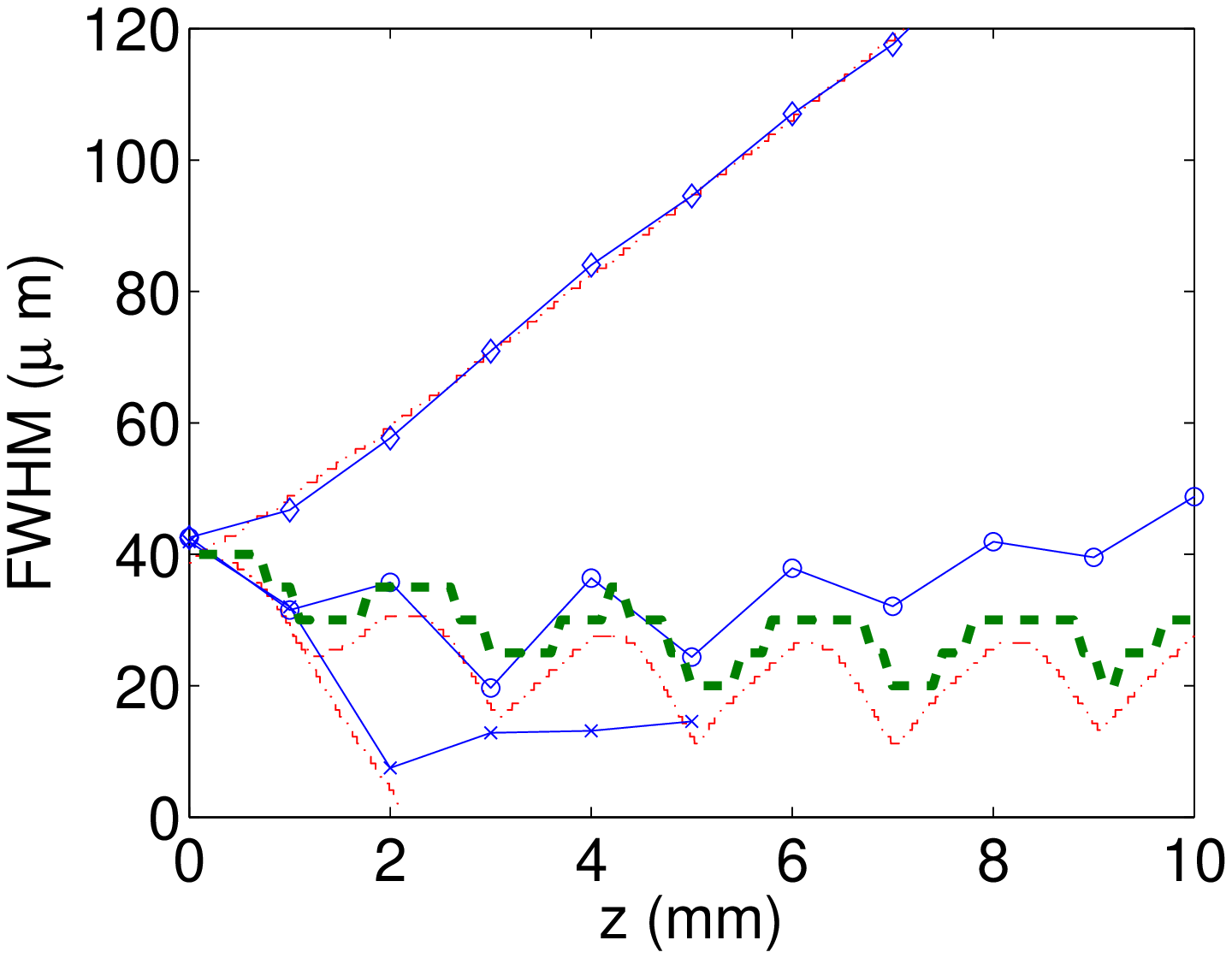}
\begin{center}
\hskip-0.2cm\includegraphics[height=5cm,width=3.75cm]{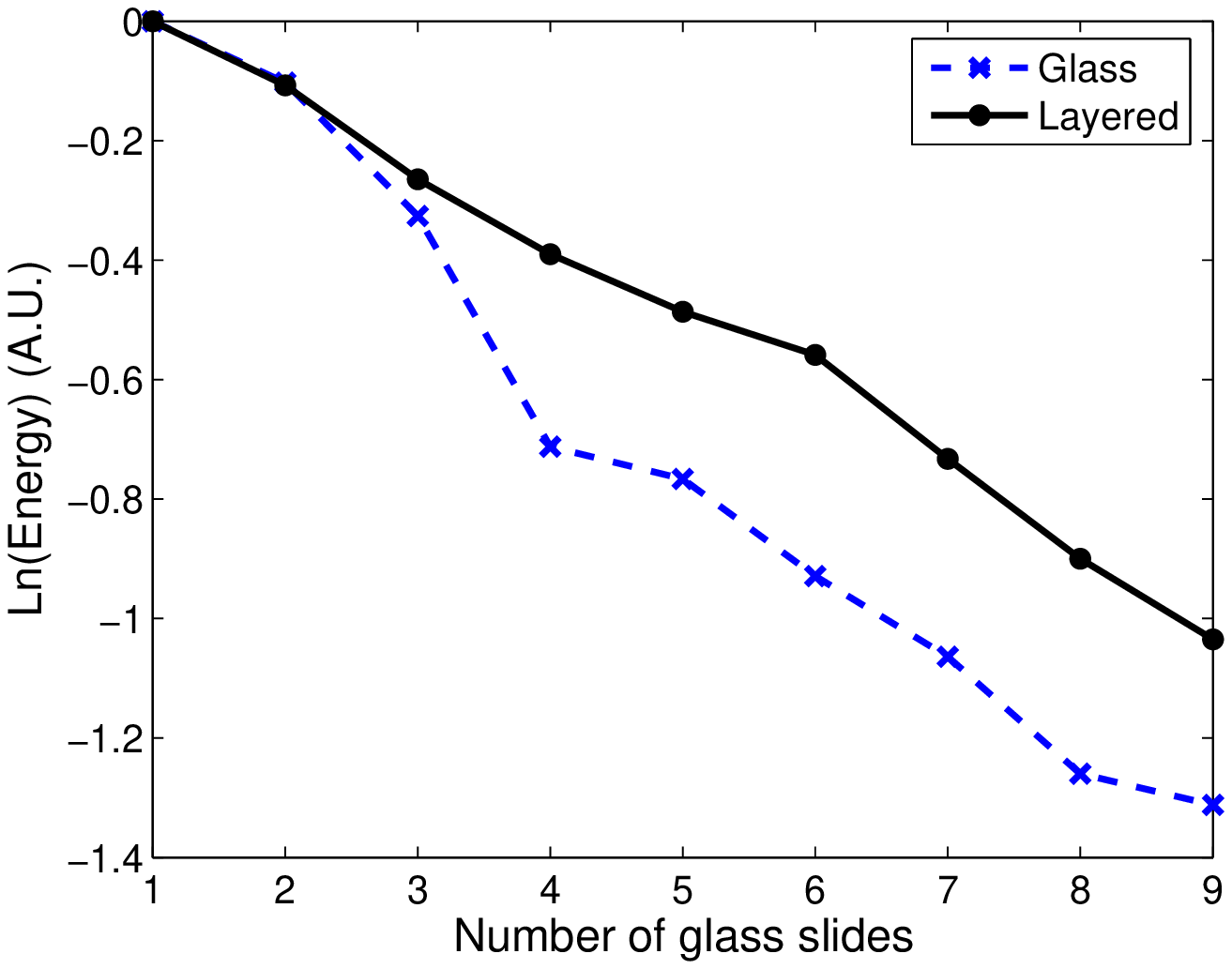}
\includegraphics[height=5cm,width=3.75cm]{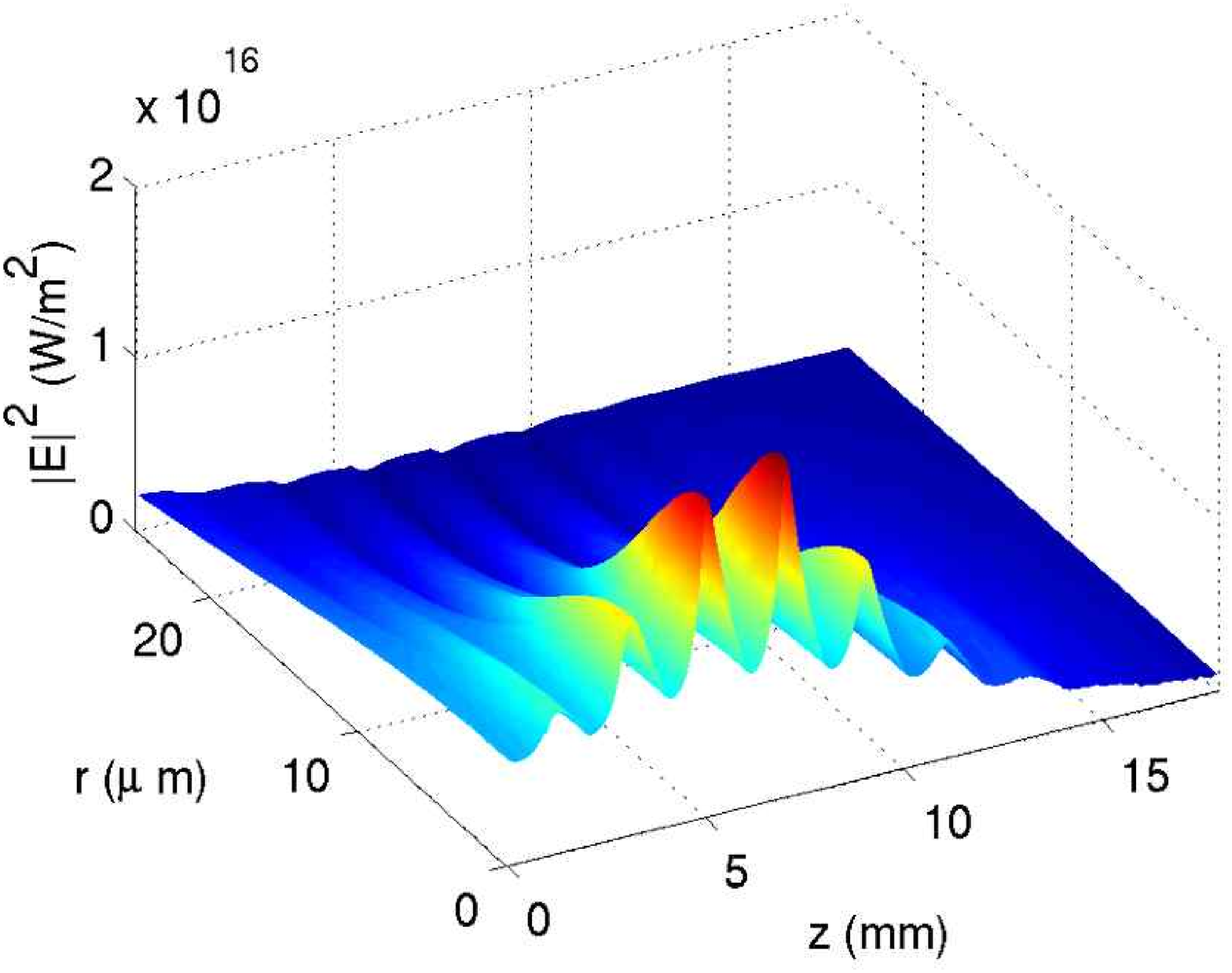}
\end{center}
\caption{(Color online) Dynamics for power $P = 5.9 P_c$.  (Top) Beam FWHM (in $\mu$m) as a function of propagation distance (in mm), for air, the layered medium, and glass. The respective experimental results are denoted by diamonds, circles, and x's (the solid curves are visual guides). The FWHM computed from Eq.~(\ref{nls}) is shown by the thin dash-dotted curve, and simulations
incorporating time-propagation are shown by the thick dashed curve.
(Bottom right) Spatial dependence of the field in Eq.~(\ref{nls}).
 (Bottom left) The (normalized) experimentally measured beam energy for glass and the layered medium as a function of the number of glass slides.} \label{energy}
\end{figure}

{\it Experimental Setup}.  We show a schematic diagram of the experimental setup in Fig.~\ref{setup}.  A Titanium:Sapphire laser amplifier system generates pulses with a duration of 160 femtoseconds and an energy of 2 mJ at $\lambda=800$ nm. The beam profile is approximately Gaussian with a measured diameter (that is, a full-width half-maximum or FWHM) of 5 mm.  The beam is attenuated using neutral density filters (ND) and then focused with a 300 mm focal length lens (L1). The nonlinear medium (NLM) is composed of 1 mm thick glass slides (Corning MicroSlides part number 2947) separated by 1 mm gaps of air. The input face of the nonlinear medium is placed 2 mm after the beam focus (the beam is diverging when it enters the nonlinear medium). At the input face of the nonlinear medium,
FWHM $\approx 43 \mu$m. After traversing the NLM, the beam is attenuated and imaged on a CCD camera (Pulnix TM-7EX) using two lenses (L2 and L3) in a 4-F configuration, with a magnification of $M = 8$.
We perform multiple experiments in which we vary the number of glass slides from one to nine. For each experiment, the beam profile is captured both at the output face of the NLM and after further propagation through 1 mm of air. The 4-F system allows us to image different planes along the propagation direction 
by changing the position of the CCD camera with respect to L3. For comparison, we also measure the propagation in glass by placing multiple glass slides together (without air gaps).

{\it Theoretical Setup}. The standard model for beam propagation in optical media, incorporating the dispersive and Kerr effects, is the NLS equation \cite{sulem}, which we adapt
to our experimental setting.
We rescale space by the wavenumber, $(\xi,\eta,\zeta)=k^{(1)}\! \ast \!(x,y,z)$ and the electric field envelope with $u=(n_2^{(1)}/n_0^{(1)})^{1/2} E$ \cite{yaron}.  (The superscript $(j)$ denotes the medium, with  $j = 1$ for glass and $j = 2$ for air.)  The physical setting is then described by
\begin{align}
    iu_\zeta & = -\frac{1}{2}{\nabla}^2_\perp u - |u|^2u \,, \quad 0 < \zeta < \tilde{l} \mbox{ (glass)}\,, \notag \\
    iu_\zeta & = -\frac{1}{2}\frac{\eta_0^{(1)}}{\eta_0^{(2)}}{\nabla}^2_\perp u - \frac{\eta_2^{(2)}}{\eta_2^{(1)}}|u|^2u \,, \quad \tilde{l} < \zeta < \tilde{L} \mbox{ (air)}\,, \label{nls}
\end{align}
where glass of scaled length $\tilde{l}$ alternates with air of scaled
length $\tilde{L}-\tilde{l}$.  Our numerical simulations consider radially symmetric solutions of (\ref{nls}), an approximation supported by the experimental data.  We compare the simulation results directly to the experiments.

To gain
analytical insight,
 we use the moment method
 \cite{montesinos} to reduce the radial NLS equation to a set of ordinary differential equations (ODEs) for its moments.
  Define
 $I_1(\zeta)
= \int |u|^2d\xi d\eta$,
$I_2(\zeta) = \int |u|^2r^2d\xi d\eta$,
$I_3(\zeta)
= i\int\left(u\frac{\partial u^*}{\partial r} - u^*\frac{\partial u}{\partial r}\right)rd\xi d\eta$,
$I_{4}^{(j)}(\zeta)
=\int\left(|\nabla u|^2 - g_j|u|^4\right)d\xi d\eta$ (with $g_1 = 1$ and $g_2 = {\eta_2^{(2)}}/{\eta_2^{(1)}}$),
and $I_5(\zeta) = \frac{1}{2}\int|u|^4d\xi d\eta$.
The quantity $I_1$ is the beam power and is conserved by the dynamics of Eq.~(\ref{nls}).  The Gaussian profile
$u =  V_0\exp\{-r^2/(2\sigma^2)\}$ gives $I_1 = V_0^2\pi\sigma^2$.  The remaining quantities are
associated with the beam width ($I_2$), momentum ($I_3$), energy ($I_4^{(j)}$), and nonlinearity ($I_5$).
Assuming
$\mbox{arg}(u) = I_3r^2/(4I_2)$,
a good approximation for $\zeta=0$ in our experiments, yields a closed set of coupled ODEs for the $I_i$. With the invariants
$Q_1^{(j)} = 2(I_4^{(j)} + g_jI_5)I_2 - I_3^2/4$ and  $Q_2 = 2\sqrt{I_2}I_5$ \cite{montesinos},
we obtain an Ermakov-Pinney (EP) \cite{ep} equation describing the dynamics of the scaled beam width $y(z) = \sqrt{I_2(\zeta)}$:
\begin{equation}
    y'' = (Q_1^{(j)} - g_jQ_2)/y^3 \equiv \beta_j/y^3 \,. \label{ep}
\end{equation}
For the initial Gaussian beam, the invariants are $Q_1^{(1)} = \pi^2\alpha^4P^2$, $Q_1^{(2)} = [\eta_0^{(1)}/\eta_0^{(2)}]Q_1^{(1)} = 1.5Q_1^{(1)}$, and $Q_2 = \pi^2\alpha^6P^3$, where $\alpha \approx 1.3556665$ and $P$ is the beam's scaled (by $P_c$) power.
In the $k$th segment of the medium, the solution is
$y = \left(A_k + B_k \zeta^2 + C_k\zeta\right)^{1/2}$,
with $A_kB_k - C_k^2 = 1$.  The initial value $y(\zeta=0)$ comes from the experimental setup, and the
coefficients $A_k, B_k$, and $C_k$ can be obtained from continuity conditions at the interfaces between the slides. Results using Eq.~(\ref{ep}) will also be directly compared with the experiments.

\begin{figure}[tbp]
\centering \includegraphics[width=7.5cm]{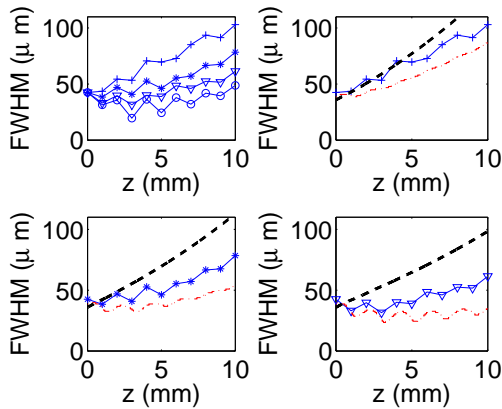} \caption{(Color online) Comparison of NM for different powers.  The top left panel shows the experimental results for $P=2.3P_c$ (pluses), $P=3.9P_c$ (stars), $P=4.9P_c$ (triangles), and $P=5.9P_c$ (circles).  We plot the FWHM as a function of the propagation distance $z$.  The individual cases of $2.3 P_c$ (top right), $3.9 P_c$ (bottom left), and $4.9P_c$ (bottom right) are compared with the PDE diagnostic (see
text), shown by the dash-dotted curves, and the ODE results
(thick dashed curves).}
\label{period}
\end{figure}

\begin{figure}[tbp]
\centering \includegraphics[width=7.5cm]{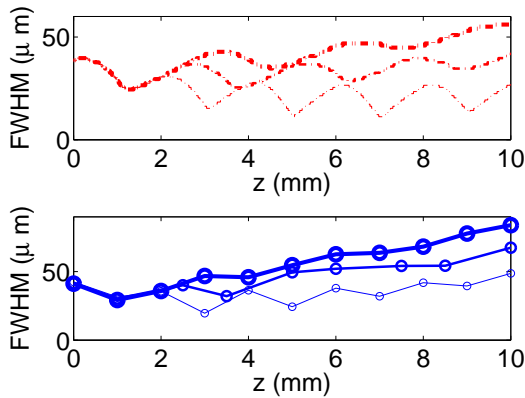} \caption{(Color online) NM for different layered media.  (Top) Numerical results for $P=5.9 P_c$ with 1 mm glass--1 mm air slide layers (thinnest dash-dotted curve), with 1 mm glass--1.5 mm air layers (intermediate thickness dash-dotted curve), and with 1 mm glass--2 mm air layers (thickest dash-dotted curve).  (Bottom) Corresponding solid curves with increasing thicknesses and symbol sizes represent the experimental results for the respective cases.}
\label{widths}
\end{figure}

{\it Results}. We applied our experimental, numerical, and analytical approaches to a variety of settings.
Figure \ref{energy} shows the measured beam diameter (FWHM) as a function of propagation distance in air, the layered medium, and glass. For the layered medium, the beam propagates through glass from 0 to 1 mm, through air from 1 mm to 2 mm, then through another layer of glass followed by air, and so on for nine periods. The pulse power is 5.9 $P_c$.  In air, the effect of the nonlinearity is practically negligible, so the beam diffracts. In glass, the beam focuses to a minimum diameter of 7.5 $\mu$m after a distance of 2 mm,
defocuses to 14 $\mu$m between 3 mm and 5 mm (the reason is discussed below), and
then starts to diverge. In
the layered medium, the beam initially focuses through the first layer and then its diameter oscillates over the first three periods with a mean of about 30 $\mu$m. The minimum beam diameter is 20 $\mu$m. The oscillations then die down and the beam starts to diverge, although with a smaller slope than either propagation in air or in glass.

Comparing the propagation through the layered nonlinear medium with the linear propagation (air), it is clear that the beam diameter is sustained over several diffraction lengths.  In both the glass and the layered medium, the beam is self-focused and stabilized. The loss of energy is what ultimately causes the beam to diverge. Their stabilizing mechanisms are different. The bottom panel of Fig.~\ref{energy} shows the total power in the beam as a function of the number of glass slides the beam has traversed for both the layered medium and glass. In the former,
we measured a total transmission of 37\% for 8 layers (88\% per glass slide),
independent of the power (linear loss). This arises mainly from reflections at the air-glass interfaces and perhaps also absorption/scattering. In glass, an additional loss appears after 2 mm when the beam reaches the minimum diameter. It has been demonstrated
for similar propagation
parameters that the stabilizing mechanism for propagation in glass is plasma formation \cite{Tzortzakis}, which occurs through multi-photon absorption that creates a negative index change and
balances the Kerr self-focusing.
The multi-photon absorption thus accounts for the additional loss measured in glass. The resulting larger beam diameter (lower intensity) prevents multi-photon absorption from becoming significant in the layered medium. The guiding in the layered medium is done purely by Kerr self-focusing and diffraction. In glass, plasma formation starts at about 2 mm, whereas here
the beam propagates through the layered medium for a much longer distance. We believe the range of stabilization in the layered medium can be greatly improved by reducing the reflection losses due to the refractive index mismatch (e.g., by using anti-reflection coating on the glass surfaces). The experimental results are in good qualitative and quantitative agreement with
the PDE simulations of Eq.~(\ref{nls}).  
In Fig.~\ref{energy}, we show that the beam's FWHM accurately follows the propagation in both air and glass. In the layered medium, the PDE
also follows the experiment qualitatively (and even quantitatively at first).  (Our simulations include the losses at each interface.)  We also used a beam propagation code to perform a full 3D simulation that includes dispersion in the temporal domain.  The results are similar to the 2D simulation, but with an improved quantitative agreement with the experiments (Fig.~\ref{energy})  
The effect of temporal dispersion is to increase the duration of the pulse, thereby decreasing the power and the strength of the self-focusing. In the simulation, the pulse broadens by approximately 15\% after 15 mm of propagation, and the shape changes from a Gaussian to a weakly multi-peaked profile. The numerical results show that temporal effects do not play a critical role in the stabilization of the beam, as shown previously for plasma stabilized filaments in glass \cite{Tzortzakis}.  In the rest of the paper, we thus compare the experimental results with the 2D simulation.


We also studied propagation through the layered medium as a function of pulse power. The top left panel of Fig.~\ref{period} shows the
evolution of the beam diameter for $P = 5.9, 4.9, 3.9, 2.3 P_c$.
We adjust the
power by placing neutral density filters before the focusing lens L1 in Fig.~\ref{setup}. There is a clear trend to stronger self-focusing with increasing power.  In each case, the beam diverges less
than for propagation in air. There are some oscillations in the beam diameter; they increase with power and eventually die down after a few periods as the power decreases. The minimum beam diameter decreases with increasing power, showing that there is no significant intensity clamping for these power values. The measured loss was the same (within experimental errors) for all power levels. The results for $2.3 P_c$ (top right panel), $3.9 P_c$ (bottom left), and $4.9 P_c$ (bottom right) are also compared with PDE and ODE numerical results.
The PDE diagnostic (shown by dash-dotted curves) is based on the definition of the original beam and
correctly follows the experiment's qualitative trends
(and even its quantitative ones for short propagation distances or weaker
powers involving quasi-linear propagation/beam divergence). The ODE
approximation (thick dashed curves) captures the weaker beam divergence for larger $P$ but is less successful with finer features (such as oscillations in the beam width). This may be attributable to the sensitive nature of the
closure approximations (especially bearing in mind the loss properties of the medium). Nevertheless,
Eq.~(\ref{ep}) provides a fair, analytically tractable approximation to the observations.

Finally, Fig.~\ref{widths} shows the effect
of changing the periodicity of the layered structure. The initial power was set to 5.9 $P_c$, and the air gaps were increased from 1 mm to 1.5 mm and then 2 mm. The thickness of the glass layers was 1 mm in each case.
Changing the periodicity effectively changes the divergence of the beam. As the period is increased, the intensity of the beam is lower when it reaches the second and subsequent glass layers, resulting in a weaker self-focusing and a ``faster'' divergence. The PDE results for the FWHM once again accurately capture the relevant trends qualitatively (and also quantitatively at first).

{\it Conclusions}. In sum, we have offered in the present work the first experimental realization of nonlinearity management (NM) in the context of optical physics using femtosecond pulse propagation in layered (glass-air) media.  We have demonstrated stabilization of the beam through NM, which can potentially provide a lossless self-guiding mechanism. We have examined the effects of different beam powers and different layered-media compositions.  We also compared these results with uniform media and (partially) accounted for the relevant loss mechanisms. The experimental results are accurately captured qualitatively (and, when appropriate, also quantitatively) by an NLS model and some of their key features can also be seen using a far more drastic but analytically tractable ODE approximation.  Interesting future experimental directions may involve the reduction of losses or the incorporation of slides of defocusing material that could lead to a complete stabilization (with stable oscillations of the beam) \cite{saito,montesinos}.

{\it Acknowledgements} We thank Michael Cross for initiating this collaboration and
acknowledge support from the DARPA Center for Optofluidic Integration (DP), the Caltech Information Science and Technology initiative (MC, MAP), and NSF-DMS and CAREER (PGK).


\end{document}